\documentclass[aps,a4paper,12pt, twocolumns]{revtex4}
\usepackage{float}
\usepackage{fancyhdr}
\usepackage{color}
\usepackage{graphicx}
\usepackage{epsfig} 
\usepackage{amssymb}
\usepackage{amsmath,amsfonts}
\usepackage{booktabs}
\usepackage{hyperref}
\usepackage{pgfplotstable}
\usepackage[footnotesize,singlelinecheck=off,justification=raggedright]{caption}
\usepackage{multirow}
\def\bra#1{\mathinner{\langle{#1}|}}
\def\ket#1{\mathinner{|{#1}\rangle}}

\def\Bra#1{\left<1>}
\DeclareMathAlphabet{\mathbbmsl}{U}{bbm}{m}{sl}
{\catcode`\|=\active\gdef\Braket#1{\left<\mathcode`\|"8000\let|\bravert {#1}\right>}}
\def\bravert{\egroup\,\vrule\,\bgroup}
\def\Tr{\mathop{\mbox{\normalfont Tr}}\nolimits}

\newcommand{\norm}[1]{\left\lVert#1\right\rVert}
\newcommand{\floor}[1]{\lfloor #1 \rfloor}
\def\cont{\mathcal{C}}
\def\r{r}

\begin{document}

\title{Sublogarithmic behaviour of the entanglement entropy in fermionic chains}

 \author{Filiberto Ares\footnote{Corresponding author.}}
\email{fares@iip.ufrn.br}
\affiliation{Departamento de F\'{\i}sica Te\'orica, Universidad de Zaragoza,
  50009 Zaragoza, Spain}
\affiliation{Centro de Astropart\'{\i}culas y F\'{\i}sica de Altas Energ\'{\i}as (CAPA),
50009, Zaragoza, Spain}
\affiliation{International Institute of Physics, UFRN, 59078-970, Natal, RN, Brazil}
\author{Jos\'e G. Esteve}
 \email{esteve@unizar.es}
 \affiliation{Departamento de F\'{\i}sica Te\'orica, Universidad de Zaragoza,
50009 Zaragoza, Spain}
\affiliation{Instituto de Biocomputaci\'on y F\'{\i}sica de Sistemas
Complejos (BIFI), 50009 Zaragoza, Spain}
\affiliation{Centro de Astropart\'{\i}culas y F\'{\i}sica de Altas Energ\'{\i}as (CAPA),
50009, Zaragoza, Spain}
  \author{Fernando Falceto}
\email{falceto@unizar.es}
 \affiliation{Departamento de F\'{\i}sica Te\'orica, Universidad de Zaragoza,
50009 Zaragoza, Spain}
\affiliation{Instituto de Biocomputaci\'on y F\'{\i}sica de Sistemas
Complejos (BIFI), 50009 Zaragoza, Spain}
\affiliation{Centro de Astropart\'{\i}culas y F\'{\i}sica de Altas Energ\'{\i}as (CAPA),
50009, Zaragoza, Spain}
  \author{Zolt\'an Zimbor\'as}
  \email{zimboras.zoltan@wigner.mta.hu}
  \affiliation{Theoretical Physics Department, Wigner Research Centre for Physics,
    P.O.Box 49 H-1525, Budapest, Hungary}
  \affiliation{MTA-BME Lend\"ulet Quantum Information Theory Research Group, Budapest, Hungary}
  \affiliation{Mathematical Institute, Budapest University of Technology and Economics, 
P.O.Box 91 H-1111, Budapest, Hungary}

%\date{\today; \quad Filename:\ {\tt\jobname.tex}}

%\begin{quote}
%{\tt  $\,$  \vskip -1.9cm         [Filename: \jobname.tex]}
%\end{quote}

\begin{abstract} 
In this paper, we discuss the possibility of unexplored
behaviours for the entanglement entropy in extended
quantum systems.
Namely, we study the R\'enyi entanglement
entropy for the ground state of long-range Kitaev chains
with slow decaying couplings.
We obtain that, under some circumstances, the entropy
grows sublogarithmically with the length of the subsystem.
Our result is based on the asymptotic behaviour
of a new class of Toeplitz determinants whose symbol
does not lie within  the application domain of the Strong Szeg\H{o} Theorem
or the Fisher-Hartwig conjecture.
  \end{abstract}

\maketitle

\section{Introduction}

In 1973 Bekenstein suggested \cite{Bekenstein} and
shortly afterwards Hawking proved \cite{Hawking} that the
entropy of a black hole is proportional to the area
of its events horizon, contrary to the usual thermal
entropy, that is proportional to the volume of the
system.
Trying to better understand the Bekenstein-Hawking
entropy, Sorkin and collaborators considered in the
pioneer work \cite{Bombelli} a real scalar field and
computed the von Neumann entanglement entropy of the
reduced density matrix of the ground state restricted
to a region of the space. They found that this entropy
is proportional to the boundary area that separates this
region from the rest of the space; i.e. similarly to the black
hole, the entropy follows an \textit{area law}.
The same property was also found later by Sredincki in \cite{Srednicki}.
This important result not only revealed that entanglement
may play a role in the black hole physics, but also indicated
that the area law may appear in other situations.

As in the example above, the entanglement entropy often follows an area law.
It is commonly admitted that this occurs in the ground
state of theories with mass gap and finite-range interactions.
This feature was indeed proved by Hastings \cite{Hastings}
for one-dimensional systems, in which the area law implies
that the entanglement entropy saturates to a constant value in the
large size limit. On the contrary, when the mass gap is zero,
the area law is violated and the entanglement entropy typically
grows with the logarithm of the length of the interval considered,
its coefficient being proportional to the central charge of
the underlying conformal field theory \cite{Holzhey, CarCal, Wolf}.

The presence of infinite-range interactions enriches the
previous picture. For example, in the fermionic ladders
studied by the authors in Ref. \cite{Ares1}, the leading
term of the ground state entanglement entropy is proportional
to the length of the interval, i.e. a \textit{volume law}.
In Ref. \cite{Vodola1}, Ercolessi and collaborators introduced
a long-range Kitaev chain with pairing couplings that decay
with a power law. These systems may exhibit a logarithmic growth
of the entropy even outside the critical points of the system
\cite{Ares3}. Moreover, as was discovered in \cite{Ares6},
the logarithmic term may be non universal and cannot be derived
from a conformal field theory.
{\color{black}Violations of the area law have been found also in 
systems without translational invariance where the couplings 
force the accumulation of singlet bonds at some points, or when there are 
random and inhomogeneous couplings \cite{Latorre,Sierra1,Dubail,Sierra2}}

A natural question is if the entanglement entropy can present
other behaviours with the size of the subsystem, apart from the
volume and the area laws or the logarithmic growth. There are some
recent works that address this issue.

In Ref. \cite{Movassagh}, Movassagh and Shor introduced a
translational invariant and local spin chain
whose ground state entanglement entropy grows with the square
root of the length of the interval. This result has been
extended to other spin chains, see e.g. Refs.
\cite{Salberger, Salberger2}. In Refs. \cite{Fannes, Farkas, Gori},
it was found that, in quadratic fermionic chains with
long-range couplings that lead to a fractal Fermi surface,
the entanglement entropy also grows with a power of the
length of the interval.

On the other hand, in Ref. \cite{Bianchini} Bianchini
et al. analyzed the entanglement entropy in non-unitary
field theories and obtained that sublogarithmic terms may
be present in the case of logarithmic conformal field theories,
see also \cite{Couvreur, Sierra3}. Some numerical studies
like \cite{Koffel} in Ising spin chains with long-range
interactions also suggest the possibility of finding
sublogarithmic terms in the entanglement entropy.

Motivated by these works, one can wonder if it is possible to
engineer a fermionic chain with long-range couplings in which
the entanglement entropy of the ground state may display a
sublogarithmic growth.

One method to compute the entanglement entropy in quadratic
fermionic chains is to express it in terms of the two-point
correlation matrix. Then, if the chain is translational invariant,
one can exploit the properties of Toeplitz and block Toeplitz
determinants and find the asymptotic expansion of the entropy
with the length of the subsystem.

The Strong Szeg\H{o} Theorem and the Fisher-Hartwig conjecture 
for Toeplitz determinants, as well as their generalisations to
the block Toeplitz case, are the two main results describing their
asymptotic behaviour. According to these results, the entanglement
entropy must grow linearly or logarithmically with the size
of the subsystem, or tend to a constant value.

However, one should take into account that these theorems have a
range of validity. The Strong Szeg\H{o} Theorem only applies
to Toeplitz matrices generated by a smooth enough, non-zero
symbol. The Fisher-Hartwig conjecture extends the former to
symbols with jump discontinuities and/or zeros.

Therefore, the fermionic chains in which the entanglement entropy
behaves in a different way would correspond to symbols that break
the hypothesis of the Strong Szeg\H{o} Theorem and the Fisher-Hartwig
conjecture. For example, a continuous, non-zero symbol 
(then the Fisher-Hartwig conjecture does not apply) but non smooth
enough according to the conditions imposed by the Strong Szeg\H{o} Theorem.

Interestingly enough, to our knowledge, the results in the literature
do not cover the asymptotic behaviour of Toeplitz determinants for
a symbol with those features.

In this paper, we shall propose a conjecture for the asymptotic
behaviour of a family of continuous symbols whose derivative diverges
at a single point, generalising the result of Fisher and Hartwig.
According to our conjecture, the leading term in the logarithm of
their Toeplitz determinant is sublogarithmic in its dimension.
The numerical checks will leave little doubt about the correctness
of this result. We will apply this result to the calculation of
the R\'enyi entanglement entropy for the ground state of
long-range Kitaev chains with pairing couplings that decay slowly
with the distance, obtaining that it grows sublogarithmically with
the length of the interval. To the best of our knowledge, this
property has not been reported before and constitutes the main
result of this work.

The paper is organized as follows. In Section \ref{sec:l_r_kitaev},
we present the physical systems of interest, the long-range Kitaev
chains, reviewing the known results for the ground state entanglement
entropy. In Section \ref{sec:sublog_conj}, we discuss the Strong Szeg\H{o}
Theorem and the Fisher-Hartwig conjecture for Toeplitz determinants
and we consider a family of symbols for which these results are not
valid. We propose for this family of symbols a new asymptotic regime
that leads to the sublogarithmic growth of the logarithm of the
determinant with its dimension. In Section \ref{sec:ent_entropy},
using this result, we extract analytically the leading term of
the R\'enyi entanglement entropy in the ground state of long-range
Kitaev chains with slow decaying pairing couplings. We finish in
Section \ref{sec:conclusions} with the conclusions. We have also
included an Appendix where we describe several results on Toeplitz
and block Toeplitz determinants that could be useful to the reader.

\section{Long-range Kitaev chains}\label{sec:l_r_kitaev}

In this paper we are interested in long range
Kitaev chains. Its Hamiltonian is given by
\begin{equation}\label{long}
  H=\sum_{m=1}^{M}\left(a_{m}^\dagger a_{m+1}+a_{m+1}^\dagger a_{m}
  +ha_{m}^\dagger a_{m}\right)+
  \sum_{m=1}^{M}\sum_{0<|l|\leq {M}/2} \gamma_l\left(a_{m}^\dagger a_{m+l}^\dagger
  -a_{m} a_{m+l}\right)
  -\frac{{M}h}{2},
\end{equation}
where $a_{m}^\dagger, a_{m}$ are the fermionic creation and annihilation operators
fulfilling the canonical anticommutation relations
\begin{equation}\label{fermionic}
 \{a_{m}^\dagger, a_n\}=\delta_{mn},\quad\{a_m^\dagger,a_n^\dagger\}=
 \{a_m,a_n\}=0,\quad m,n=1,\dots,{M},
\end{equation}
and we consider periodic boundary conditions, $a_{{M}+m}=a_{m}$.
The hopping terms only couple nearest neighbours but the pairing
$\gamma_l$ extends to the whole chain and satisfies $\gamma_l=-\gamma_{-l}$.
In order to keep certain sense of locality we shall assume that
they attenuate at infinity, i.e. $\gamma_l\to 0$ when $l\to\pm\infty$.

This is a free, translational invariant system whose elementary
excitations are characterized by its momentum $k$. The
dispersion relation tells us the energy associated to each mode.
In the thermodynamic limit and denoting
by $\theta=2\pi k/{M}$ it is
\begin{equation}\label{dispersion}
  \omega(\theta)=\sqrt{(h+2\cos\theta)^2+|G(\theta)|^2},
\end{equation}
where
\begin{equation}\label{funG}
  G(\theta)=\sum_{l=1}^\infty\left({\rm e}^{{\rm i}\theta l}
  -{\rm e}^{-{\rm i}\theta l}\right)\gamma_l
\end{equation}

An example of this system that has been most thoroughly studied corresponds
to
$$
{\color{black}\gamma_l= {\rm sign}(l)\,|l|^{-\delta},\qquad \delta>0, \qquad  l\ne 0.}$$
Therefore, the Hamiltonian takes the form
\begin{equation}\label{lrk}
  H_{\rm LRK}=\sum_{m=1}^{M}\left(a_{m}^\dagger a_{m+1}+a_{m+1}^\dagger a_{m}
  +ha_{m}^\dagger a_{m}\right)+
  \sum_{m=1}^{M}\sum_{0<|l|\leq {M}/2}l|l|^{-\delta-1}\left(a_{m}^\dagger a_{m+l}^\dagger
  -a_{m} a_{m+l}\right)
  -\frac{{M}h}{2}.
\end{equation}

This model was introduced a few years ago
in the pioneering paper \cite{Vodola1}. 

Since then it has been the object of an intense study. It is 
very useful to analyse the effects of long-range interactions. Here we shall 
restrict to the entanglement entropy but different authors have used the model
to investigate the form of the correlations \cite{Vodola1, Vodola2, Cirac4},  
the breaking of conformal symmetry \cite{Lepori}, the propagation of information 
\cite{Regemortel}, the behaviour out of equilibrium \cite{Lepori2} 
or the occurrence and structure of topological phases 
\cite{Patrick, Lepori3, Alecce, Viyuela, Cats, Sedlmayr}.

As we just mentioned, in this paper we will be interested in the R\'enyi
entanglement entropy for a subsystem
$X$ in the ground state $\ket{{\rm GS}}$
of the chain. It is defined by
$$S_{\alpha,X}=\frac1{1-\alpha}\log\Tr_X(\rho_X^\alpha),$$
where $\rho_X=\Tr_Y\ket{{\rm GS}}\bra{{\rm GS}}$ is the reduced density
matrix obtained by tracing out, in the pure ground state,
the degrees of freedom of the subsystem $Y$ complementary to $X$.
In the limit $\alpha\to 1$ we recover the von Neumann entropy.

In the original paper \cite{Vodola1}
the authors compute numerically the scaling behaviour
of the entanglement entropy as a function of the
dumping exponent $\delta$ for $\delta\not=1$.
The interest of the model resides in the fact that it violates
the standard connection with conformal field theories.

In fact, according to the latter, for a one dimensional,
critical, local theory the R\'enyi entanglement entropy of a subsystem
$X$ grows logarithmically with its size $|X|$  with a coefficient
proportional to the central charge of the theory ${\rm c}$. Namely
\begin{equation}\label{CFT}
 S_{\alpha, X}\sim\frac{\alpha+1}{6\alpha}{\rm c}\log |X|.
\end{equation}

Also, in a free theory like the one at hand, the central charge
is the number of massless bosonic particles or
one half of the fermionic ones.
In concrete terms, in our fermionic spinless system,
the central charge corresponds to one half 
of the number of zeros in the dispersion relation.

In order to apply the previous results to (\ref{lrk})
we must compute the dispersion relation, i.e. the energy
of the single particle eigenstate of momentum $k$.
Particularizing (\ref{dispersion}) to the case (\ref{lrk})
we have that, in the thermodynamic limit,
\begin{equation}\label{dispersion_lrk}
  \omega_{\rm LRK}(\theta)=\sqrt{(h+2\cos\theta)^2+|G_{\rm LRK}(\theta)|^2},
\end{equation}
where 
\begin{equation*}
  G_{\rm LRK}(\theta)={\rm Li}_\delta({\rm e}^{{\rm i}\theta})-
  {\rm Li}_\delta({\rm e}^{-{\rm i}\theta}),
\end{equation*}
and
${\rm Li}_\delta$ 
stands for the polylogarithm of order $\delta$ \cite{NIST},
\begin{equation*}
 {\rm Li}_\delta(z)=\sum_{l=1}^\infty\frac{z^l}{l^{\delta}}.
\end{equation*}
This is analytic for $|z|<1$ and extends, for $\delta>0$, to a
multivalued function in the whole complex plane excluding, in some cases,
$z=1$. In fact, ${\rm Li}_\delta$ has a finite limit at $z=1$
for $\delta>1$ while it 
diverges at this point for $\delta\leq1$.
For $\delta=1$ the polylogarithm function reduces to the logarithm,
namely ${\rm Li}_1(z)=-\log(1-z)$.

{\color{black}
As we mentioned before, massless excitations are associated to
the zeros of the dispersion relation (\ref{dispersion_lrk}). It
only vanishes for $h=2$ at $\theta_F=\pi$ or for $h=-2$ and $\theta_F=0$
\footnote{
{\color{black}Note that near these points the energy behaves linearly 
  with $|\theta-\theta_F|$, as in a massless relativistic dispersion
  relation.}}.}
Therefore, according to the general theory (for local systems)
one could expect that the entanglement entropy in the vacuum
state goes to a constant for $h\not=\pm 2$ in the large $|X|$
limit and it behaves like
\begin{equation*}
 S_{\alpha, X}\sim\frac{\alpha+1}{6\alpha}\frac12\log |X|.
\end{equation*}
for $h=\pm2$.

In \cite{Ares3} and \cite{Ares6}, using some novel results derived
by the authors on the asymptotic behaviour for block Toeplitz
determinants with discontinuous symbols, we have been able to compute
the large $|X|$ limit of the entanglement entropy for the vacuum state
of the long range Kitaev chain (\ref{lrk}). The results, that we will
outline below, do not follow the previous guess and rather confirm
the numerical research of \cite{Vodola1}.

We first consider the case $\delta\not=1$ that was the subject of refs.
\cite{Vodola1, Ares3}. In this case we can show that the entanglement
entropy follows the general form
  \begin{equation*}
    S_{\alpha, X}\sim\frac{\alpha+1}{6\alpha}{\rm c}\log |X|,
  \end{equation*}
but contrary to what is expected from conformal field theory
($\rm c=1/2$ for $h=\pm2$ and $\rm c=0$ otherwise) we have the
following:
  \begin{equation*}
    {\rm c}=
    \begin{cases}
      0&\mbox{for } \delta>1  \mbox{ and }  h\not=\pm2 \cr
      1/2&\mbox{for } \delta<1 \mbox{ and } h\not=2  \mbox{ or } \delta>1 \mbox{ and } h=\pm2\cr
      1&\mbox{for } \delta<1 \mbox{ and } h=2.
    \end{cases}
  \end{equation*}

  Of course, the explanation for this deviation from the predicted result
  lays in the fact that the theory is non local (it has non zero coupling at
  all distances) and therefore the premises of the conformal field theory
  prediction are not met.

  Anyhow, a more interesting result is obtained for $\delta=1$. This case is
  technically more complex and it has been discussed thoroughly in \cite{Ares6}.
  In this situation the R\'enyi entanglement entropy does not even follow
  the general expression (\ref{CFT}), but rather it behaves like
  \begin{equation*}
    S_{\alpha, X}=B_{\alpha}\log |X|+O(1),
  \end{equation*}
  where
  \begin{equation}\label{b0}
B_{\alpha}=
\frac{2}{\pi^2}\int_{\cos \xi}^1
\frac{{\rm d} f_\alpha(\lambda)}{{\rm d}\lambda}
\log\frac{\sqrt{1-\lambda^2}}{\sqrt{\lambda^2-\cos^2\xi}+\sin\xi}
         {\rm d}\lambda,
  \end{equation}
  with
  $$\cos\xi=\frac{h+2}{\sqrt{(h+2)^2+\pi^2}},\quad
  \sin\xi=\frac{\pi}{\sqrt{(h+2)^2+\pi^2}}.
  $$
  and\begin{equation}\label{falpha}
  f_\alpha(x)=\frac{1}{1-\alpha}\log\left[\left(\frac{1+x}{2}\right)^\alpha
    +\left(\frac{1-x}{2}\right)^\alpha\right].
  \end{equation}
  
 {\color{black}
  %%To our knowledge, this result is the first instance in which the  coefficient
  Therefore, the coefficient
  of the logarithmic term in the asymptotic behaviour of the R\'enyi
  entanglement entropy does not depend on $\alpha$ as in (\ref{CFT}),
  loosing any connection with conformal field theory. Similar type of
  results can be found in \cite{Refael,Laflorencie}.}

  The goal of the next sections is to go even further in exploring possible
  anomalous asymptotic behaviours for the entanglement entropy.
  In fact, we will show that it is possible to choose the pairing $\gamma_l$
  in (\ref{long}) so that the entropy  grows like a fractional power of
  the logarithm of the size of the subsystem $X$.

  \section{Sublogarithmic behaviour of Toeplitz determinants}\label{sec:sublog_conj}

  As it has been discussed previously in a number of papers,
  see e.g. \cite{Jin, Its, Its2, Kadar, Ares1, Ares3, Ares6, Kormos},
  and will be recalled in the next section, the entanglement
  entropy of the free fermionic chain can be derived from the
  computation of a Toeplitz determinant. Therefore, before
  addressing the problem mentioned at the end of the previous
  section we have to establish some new results on the asymptotic
  behaviour of Toeplitz determinants in the following paragraphs.

  The history of Toeplitz operators and Toeplitz determinants
  is an interesting example of crossbreading between physics
  and mathematics. It originated from a branch of pure abstract
  mathematics that very soon proved to be essential for physics:
  the Hilbert space. 

  Toeplitz operators were introduced in 1907 by Otto Toeplitz
  \cite{Toeplitz}, at G\"ottingen at that time, to provide
  concrete examples of the abstract theory that David Hilbert
  was developping there. We will be interested in the asymptotic
  behaviour of Toeplitz determinants as defined below.

  For a complex function $f$ on the unit circle $S^1$ with
  $f\in L^1(S^1)$, we shall denote by $T_N[f]$ the Toeplitz
  matrix with symbol $f$ and dimension $N\times N$. Its
  entries $(T_N[f])_{nm}=f_{n-m}$ are given by the Fourier
  coefficients of the function $f$,
  \begin{equation*}
   f_k=\frac{1}{2\pi}\int_{-\pi}^\pi f(\theta){\rm e}^{{\rm i}\theta k}{\rm d}\theta.
  \end{equation*}

The first result on the large $N$ limit of its determinant
$D_N[f]\equiv \det T_N[f]$ was conjectured by Polya \cite{Polya}
and proved by Szeg\H{o} in \cite{Szego}. It states the following:

\textbf{First Szeg\H{o} theorem} (1915)

\textit{Assume that $f\in L^1(S^1)$ is not vanishing and has zero winding around
zero, then its Toeplitz determinant verifies 
\begin{equation}\label{szego}
  \log D_N[f]=\frac{N}{2\pi}
  \int_{-\pi}^\pi \log f(\theta)
       {\rm d}\theta+o(N).
\end{equation}
}

This result was not improved until the early 50's motivated by
Kauffman and Onsager discoveries in statistical physics \cite{Deift0}.
Actually in 1949 they established that the magnetization of the
classical Ising model in two dimensions below critical temperature
could be obtained from the computation of a particular Toeplitz
determinant \cite{Kaufman}. But Szeg\H{o}'s theorem on the asymptotic limit of the
determinant gives a trivial result in their case, and the answer was in
the $o(N)$ correction of (\ref{szego}) which is not determined
by the theorem.

Using an alternative derivation of the magnetization,
Onsager apparently formulated a conjecture for the subleading term,
but he did not publish neither the conjecture nor the result on the
magnetization.

While Onsager was trying to prove his guess, Szeg\H{o} was aware of
the problem and soon after he came out with the proof.
Today it is known as the \textbf{Strong Szeg\H{o} Theorem} (1952)
that can be stated as follows \cite{Szego2, Ibraginov}:

\textit{Assume that the symbol $f\in L^1(S^1)$ is not vanishing, has zero winding 
number around zero and it is well-behaved in the sense that its Fourier
modes $f_k$ satisfy 
\begin{equation}\label{well-behaviour}
  \sum_{k=-\infty}^\infty |f_k|+
  \sum_{k=-\infty}^\infty |k| |f_k|^2<\infty,
\end{equation}
then the correction in (\ref{szego}) is finite 
in the limit $N\to\infty$,
\begin{equation}\label{sst}
 \log D_N[f]=\frac{N}{2\pi}\int_{-\pi}^{\pi} \log f(\theta){\rm d}\theta+\sum_{k=1}^\infty k s_k s_{-k}+o(1),
\end{equation}
where the $s_k$'s in the sum are the Fourier modes of $\log f(\theta)$,
\begin{equation*}
 s_k=\frac{1}{2\pi}\int_{-\pi}^\pi 
 \log f(\theta){\rm e}^{{\rm i}\theta k}{\rm d}\theta.
\end{equation*}
}

A little later it became clear that statistical physicists needed
to know the asymptotic behaviour for Toeplitz determinants with
more general symbols. In fact as it was shown in 1963 by Lenard
\cite{Lenard} the one particle density matrix of impenetrable
bosons can be expressed as a Toeplitz determinant with zeros in
the symbol which are excluded in Szeg\H{o}'s theorems. Also in
the Ising model at critical temperature the magnetization derives
from a Toeplitz determinant for a symbol with jump discontinuities.
In 1966 Wu \cite{Wu} was able to derive the asymptotic behaviour
of these determinants.

Putting these previous results together Fisher and Hartwig arrived
in 1968 at their remarkable conjecture published in \cite{Fisher}
and proven in full generality in 1979 by Basor in \cite{Basor}.

Here we shall discuss a particular version of the conjecture that only
allows for discontinuities of the symbol (not zeros)
as these are the kind of Toeplitz determinants that appear in
connection with the entanglement entropy.

Accordingly, consider that the symbol $f$ has jump discontinuities at 
$\theta_1,\dots,\theta_R$. The Fisher-Hartwig conjecture for this case says
that the discontinuities contribute to $\log D_N[f]$ with a logarithmic
term, whose coefficient only depends on the lateral limits 
$f_r^\pm = \lim_{\theta\to\theta_r^\pm} f(\theta)$, $r=1,\dots, R$. 
That is,
\begin{equation}\label{f-h}
 \log D_N[f]=N s_0+
 \frac{\log N}{4\pi^2}
 \sum_{r=1}^R
 \left(\log\frac{f_r^+}{f_r^-}\right)^2
 +\log E[f]+o(1).
\end{equation}

The reasoning followed by Fisher and Hartwig
in \cite{Fisher} to deduce the expansion (\ref{f-h})
is as follows.

For simplicity, and as they precisely did, we take the symbol
\begin{equation}\label{f_0}
 f^{(0)}(\theta)=
 {\rm e}^{\beta(\theta-\pi{\rm sign}(\theta))},
 \quad \theta\in[-\pi,\pi),
\end{equation}
that only has a discontinuity at $\theta=0$. 

The Fourier coefficients of $\log f^{(0)}$ are 
\begin{equation*}
 s_0^{(0)}=0,\quad \mbox{and}\quad 
 s_k^{(0)}=\frac{\beta}{{\rm i}k}, \quad \mbox{for}\quad
 k\neq 0.
\end{equation*}
Then if we apply to this symbol the Strong Szeg\"o Theorem
(\ref{sst}) we obtain the Harmonic series
\begin{equation*}
 \log D_N[f^{(0)}]=Ns_0^{(0)}+\sum_{k=1}^\infty ks_k^{(0)} s_{-k}^{(0)}+o(1)
 =\sum_{k=1}^\infty \frac{\beta^2}{k}+o(1),
\end{equation*}
that diverges logarithmically. 

Let us suppose that $\log D_N[f^{(0)}]$ can be obtained truncating 
the series $\sum_{k=1}^\infty ks_k^{(0)}s_{-k}^{(0)}$ at some 
$k=\floor{N \Lambda_0}$, with $\Lambda_0$ a positive real number
(here $\floor{t}$ means integer part of 
$t$). Then we find
\begin{equation}\label{f-h_trun}
 \log D_N[f^{(0)}]=\sum_{k=1}^{\floor{N\Lambda_0}}ks_k^{(0)}s_{-k}^{(0)}=
 \sum_{k=1}^{\floor{N\Lambda_0}}\frac{\beta^2}{k}=
 \beta^2\log(N\Lambda_0)+\beta^2\gamma_{\rm E}+o(1), 
\end{equation}
where $\gamma_{\rm E}$ is the Euler-Mascheroni constant.

Notice that this truncation gives precisely the Fisher-Hartwig 
expansion (\ref{f-h}) for the symbol $f^{(0)}$. In fact, since
$f^{(0)}(\theta)$ presents a single discontinuity ($R=1$) with
lateral limits ${\rm e}^{\pm \beta\pi}$, the expression in (\ref{f-h})
particularises as
\begin{equation}\label{f-h2} 
 \log D_N[f^{(0)}]=\beta^2\log N+\log E[f^{(0)}]+o(1).
\end{equation}
Comparing (\ref{f-h2}) with (\ref{f-h_trun}) we can conclude that
\begin{equation}\label{cutoff_0}
 \log E[f^{(0)}]=\beta^{2}(\log \Lambda_0+\gamma_{\rm E}).
\end{equation}
Fisher and Hartwig were able to fix the constant term $E[f^{(0)}]$ 
and, therefore, the cutoff parameter $\Lambda_0$ because they 
realised that the Toeplitz matrix with symbol $f^{(0)}$ is also a 
Cauchy matrix \footnote{A matrix is of Cauchy type if its entries 
are of the form $C_{nm}=(X_n-Y_m)^{-1}$  with $X_n-Y_m\neq 0$, 
and $X_n, Y_m\in \mathbb{C}$ where $n,m \in \mathbb{N}$.}. 
Using the properties of the determinants of Cauchy matrices they 
determined that
\begin{equation*}
  E[f^{(0)}]={\rm G}(1+{\rm i}\beta){\rm G}(1-\rm{i}\beta)
\end{equation*}
where ${\rm G}(z)$ is the Barnes G-function. Hence we have 
\begin{equation}\label{Lambda_0}
 \log \Lambda_0=2\beta^{-2}\log|{\rm G}(1+{\rm i}\beta)|-\gamma_{\rm E}.
\end{equation}
The same reasoning can be applied to a general symbol 
with $R$ discontinuities. The asymptotic behaviour of its 
determinant predicted by (\ref{f-h}) can be deduced from 
the Strong Szeg\H{o} theorem (\ref{sst}) truncating the 
divergent terms in the series $\sum_{k=1}^\infty k s_k s_{-k}$.

We want to go one step further in this generalization process.
For that consider the symbol
\begin{equation}\label{mysymbol}
 \log f^{(\nu)}(\theta)= 
 \beta\frac{\theta-\pi\,{\rm sign}(\theta)}
 {\left(-\log\frac{|\theta|}{2\pi}\right)^\nu},
 \quad \theta\in[-\pi,\pi), 
\end{equation}
with $\nu\geq 0$. 
% $\beta<1$ and% 

It is clear that for $\nu>0$, $f^{(\nu)}$ is
continuous but not differentiable at $\theta=0$.
The Fourier coefficients of  $\log f^{(\nu)}$ are
\begin{equation}\label{sk}
 s_{k}^{(\nu)}=\frac{\beta}{{\rm i}k(\log|2\pi k|)^\nu}
 \left[1+O\left(\frac{1}{\log|k|}\right)\right].
\end{equation}
and, therefore, the series
\begin{equation}\label{series}
 \sum_{k=1}^\infty k s_k^{(\nu)}s_{-k}^{(\nu)},
\end{equation}
diverges for $\nu\leq1/2$.

Now we proceed in a way similar to that of Fisher and Hartwig and establish
the following conjecture
\begin{equation*} 
 \log D_N[f^{(\nu)}]=N s_0^{(\nu)}+\sum_{k=1}^{\floor{N\Lambda_\nu}}
 k s_{k}^{(\nu)}s_{-k}^{(\nu)}+o(1).
\end{equation*}
where $\Lambda_\nu$ is a constant to be determined.

Since $\log f^{(\nu)}(\theta)$ is a real odd function,
its Fourier coefficients are imaginary and such that
$s_{-k}^{(\nu)}=-s_{k}^{(\nu)}$. Therefore  $s_0^{(\nu)}=0$ 
and the term linear in $N$ in the expression above cancels.
Hence 
\begin{equation}\label{sublog_conj}
 \log D_N[f^{(\nu)}]= \sum_{k=1}^{\floor{N\Lambda_\nu}}
 k |s_{k}^{(\nu)}|^2+o(1).
\end{equation}
One easily sees that the conjecture predicts a sublogarithmic growth
of $\log D_N[f^{(\nu)}]$ with the dimension $N$. In fact, if we consider
the asymptotic behaviour (\ref{sk}) of $s_k^{(\nu)}$ we have for $0<\nu<1/2$
\begin{equation*}
\log D_N[f^{(\nu)}]=\frac{\beta^2}{1-2\nu}(\log N)^{1-2\nu}+ o(1),
\end{equation*}
while for $\nu=1/2$
\begin{equation*}
 \log D_N[f^{(1/2)}]=\beta^2\log\log N+ o(1).
\end{equation*}

\begin{figure}[H]
  \centering
    \includegraphics[width=0.9\textwidth]{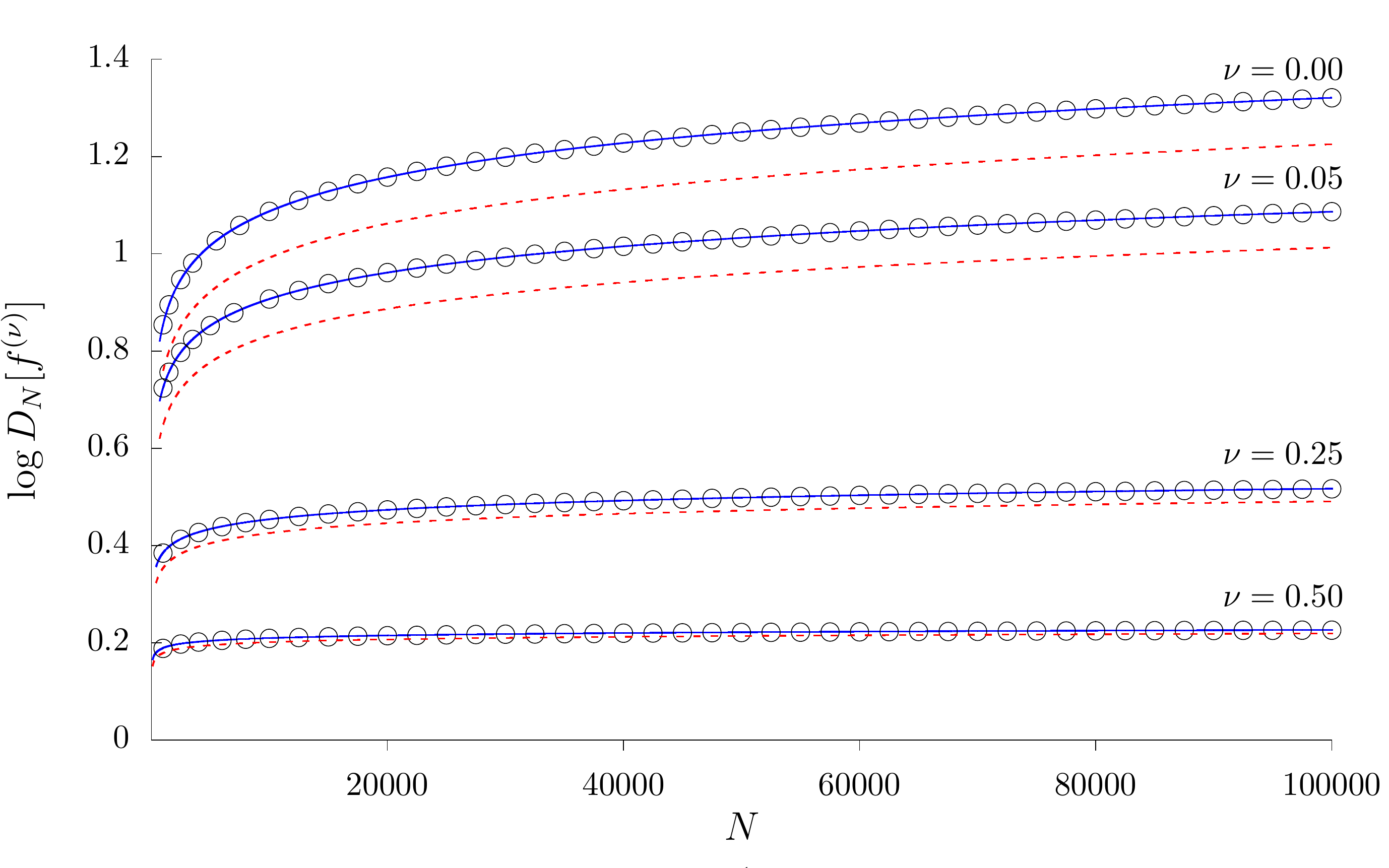}
    \caption{Logarithm of the Toeplitz determinant with symbol
      $f^{(\nu)}$, defined in (\ref{mysymbol}), plotted against
      the dimension $N$. The dots correspond to the numerical
      values obtained using the supercomputer
      \textit{Memento} for different exponents $\nu$ and $\beta=1/\pi$.
      When $0\leq\nu\leq 1/2$, $f^{(\nu)}$ does not satisfy the 
      well-behavedness condition (\ref{well-behaviour}) of the Strong
      Szeg\H{o} Theorem. The solid lines represent the conjecture
      proposed in (\ref{sublog_conj}) for the asymptotic
      behaviour of $\log D_N[f^{(\nu)}]$,
      $\sum_{k=1}^{\floor{N\Lambda_\nu}} k |s_k^{(\nu)}|^2$,
      with $\Lambda_\nu$ as given in Table \ref{tab1}. 
      For $\nu=0$ the symbol is discontinuous and we can apply
      the Fisher-Hartwig conjecture. In this case $\Lambda_0$ can
      be directly calculated using (\ref{Lambda_0}).
      The dashed lines correspond to the sum $\sum_{k=1}^N k |s_k^{(\nu)}|^2$,
      that is considering $\Lambda_\nu=1$.
    }
  \label{sublog_todas}
   \end{figure}

We can confront our conjecture (\ref{sublog_conj}) with the determinant
evaluated numerically for large values of $N$, up to $10^5$.
The results are summarized in Fig. \ref{sublog_todas}. We have
considered four values of $\nu$ ($\nu=0, 0.05, 0.25 \mbox{ and } 0.50$)
and $\beta=1/\pi$.

The values of  $\Lambda_\nu$ that give the best fit are given in Table \ref{tab1}.

\begin{table}[H]
   \centering
   \begin{tabular}{cccccc}
   \toprule
   $\nu$ & & 0.00 & 0.05 & 0.25 & 0.50 \\
   %\hline
   $\Lambda_\nu$ & & 2.566 & 2.599 & 2.659 & 2.660    \\
  \bottomrule
  \end{tabular}
   \caption{Values of $\Lambda_\nu$ that give the best fit of the
     curve $\sum_{k=1}^{\floor{N\Lambda_\nu}} k |s_k^{(\nu)}|^2$ to the
     numerical points of Fig. \ref{sublog_todas} for the different
     values of $\nu$ and $\beta=1/\pi$. The Fourier coefficients
     $s_k^{(\nu)}$ have been computed numerically.
     The value of $\Lambda_\nu$ for the case $\nu=0.00$ has been calculated
     using the expression (\ref{Lambda_0}). }
  \label{tab1}
\end{table}

The accuracy of the fit is remarkable keeping in mind
that it has been obtained by adjusting just one free parameter.
For us this is a very strong indication that our conjecture is correct.

In the following section we will apply this result to
compute the entanglement entropy
for a Kitaev chain with long range interaction.

\section{Application to the entanglement entropy}\label{sec:ent_entropy}

 Here we will apply the results of the previous sections to study 
 the asymptotic behaviour of the R\'enyi entanglement entropy
 of the following fermionic 
 chain,
 \begin{equation}\label{ham_log}
  H_{{\rm log}}=\sum_{m=1}^{M}\left(h a_m^\dagger a_m 
  + a_m^\dagger a_{m+1}
  +a_{m+1}^\dagger a_m\right)+
  \sum_{m=1}^{M}\sum_{0<|l|\leq {M}/2}\frac{1}{l(\log(A |l|))^\nu}
  \left(a_m^\dagger a_{m+l}^\dagger - a_m a_{m+l}\right),
 \end{equation}
where the pairing couplings decay logarithmically
with an exponent $0\leq \nu\leq 1/2$ and $A>1$. 
We shall consider $h\neq \pm 2$. Note that the value
$\nu=0$ corresponds  to the Long-Range Kitaev chain introduced
in Section \ref{sec:l_r_kitaev}.

In order to derive the asymptotic behaviour of the
entanglement entropy for this system we shall need
two technical results on block Toeplitz determinants
that we borrow from different authors. For the purpose
of keeping the paper readable yet self contained, we
present these theorems, together with the necessary
definitions, in the Appendix.

According to the general discussion of Section
\ref{sec:l_r_kitaev}, in the limit ${M}\to\infty$,
the dispersion relation of $H_{\rm log}$ is
\begin{equation*}
 \omega_{\rm log}(\theta)=
 \sqrt{F(\theta)^2+|G_\nu(\theta)|^2},
\end{equation*}
where
\begin{equation*}
 F(\theta)=h+2\cos(\theta),
\end{equation*}
and
\begin{equation*}
 G_\nu(\theta)=\sum_{l=1}^\infty \frac{1}{l(\log(A|l|))^\nu}
 \left({\rm e}^{{\rm i}\theta l}-{\rm e}^{-{\rm i}\theta l}\right).
\end{equation*}
Notice that the Fourier coefficients of $G_\nu$,
$1/(l(\log A |l|)^\nu)$, are similar to the leading term
in the expansion (\ref{sk}) for the Fourier coefficients 
$s_k^{(\nu)}$ of the function $\log f^{(\nu)}$ introduced in 
(\ref{mysymbol}). Hence the behaviour of $G_\nu(\theta)$
in a neighbourhood of $\theta=0$ is equal to that of
$\log f^{(\nu)}(\theta)$. We have seen that the divergence
in the derivative of $\log f^{(\nu)} (\theta)$ at $\theta=0$
violates the well-behaviour condition (\ref{well-behaviour})
of the Strong Szeg\H{o} Theorem when $0< \nu\leq 1/2$.
This divergence gives rise to a sublogarithmic growth of the
Toeplitz determinant with symbol $f^{(\nu)}$. Since $G_\nu(\theta)$
displays the same behaviour at $\theta=0$, we expect that
this affects the asymptotics of the ground state entanglement entropy.
We will show that this is indeed the case but we warn the reader that the
derivation, that we outline in the rest of the section,
is by no means immediate.

\begin{figure}[h]
  \centering
    \resizebox{12cm}{4cm}{\includegraphics{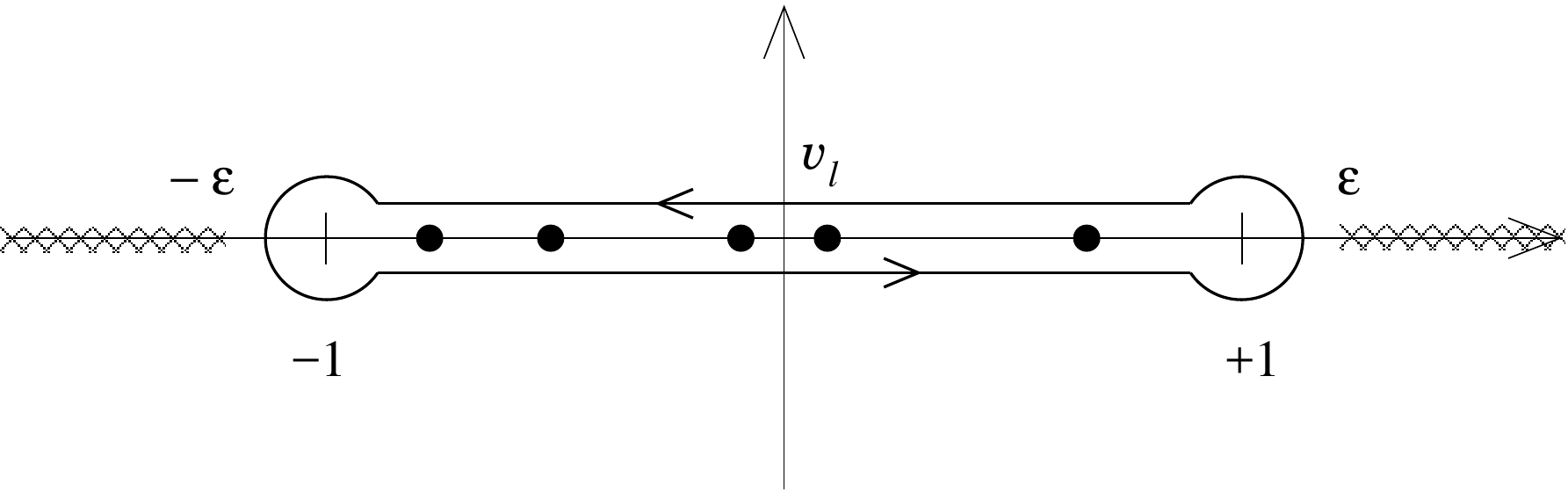}} 
    \caption{Contour of integration, cuts and poles for the computation of 
      $S_{\alpha, X}$ in (\ref{ent_contour}). The contour surrounds the eigenvalues
      $v_l$ of $V_X$, all of them lying on the real interval $[-1,1]$. The branch
      cuts for the function $f_\alpha$ extend to $\pm\infty$.}
  \label{contorno0}
\end{figure}

Following Ref. \cite{Jin} we will compute the R\'enyi
entanglement entropy for a subsystem $X$ of this chain
using
\begin{equation}\label{ent_contour}
 S_{\alpha,X}=\frac{1}{4\pi{\rm i}}
 \lim_{\varepsilon\to 1^+}
 \oint_{\cont}  f_\alpha(\lambda/\varepsilon)
   \frac{{\rm d}}{{\rm d}\lambda}
 \log D_X(\lambda){\rm d}\lambda,
\end{equation}
where $\cont$ is the integration contour 
represented in Fig. \ref{contorno0} and  
$D_X(\lambda)=\det(\lambda I- V_X)$ with
$V_X$ the restriction of the ground state 
correlation matrix to $X$,
\begin{eqnarray}\label{correl}
  (V_X)_{nm}&=&2 \left\langle 
  \left(\begin{array}{c}a_n \\ a_n^\dagger \end{array}
  \right)(a_m^\dagger,a_m)\right\rangle-\delta_{nm} I,
  \quad n,m=1,\dots, |X|.
\end{eqnarray}

If the subsystem is a single interval of length 
$|X|$ then, in the limit $M\to\infty$, $V_X$ is
a block Toeplitz matrix with entries
\begin{equation*}
(V_X)_{nm}=\frac{1}{2\pi}\int_{-\pi}^\pi
\mathcal{G}(\theta){\rm e}^{{\rm i}\theta(n-m)}
{\rm d}\theta,\quad n,m=1,\dots, |X|,
\end{equation*}
and symbol $\mathcal{G}(\theta)$, the 
$2\times 2$ matrix
\begin{equation*}
 \mathcal{G}(\theta)=\frac{1}{\omega_{\rm log}(\theta)}
 \left(\begin{array}{cc}
        F(\theta) & G_\nu(\theta) \\
        -G_\nu(\theta) & -F(\theta)
       \end{array}\right).
\end{equation*}
Our goal is to derive the leading term
in the asymptotic expansion of the block
Toeplitz determinant $D_X(\lambda)$ with 
symbol $\mathcal{G}_\lambda=\lambda I-\mathcal{G}$.
The techniques developed in the previous
section are valid for Toeplitz matrices,
that is, when the symbol is a scalar function. 
Therefore, the first step in order to analyse 
the asymptotic behaviour of $D_X(\lambda)$
is to express it in terms of Toeplitz 
determinants.

For this purpose, let us perform the 
following change of basis,
\begin{equation*}
 \mathcal{G}_\lambda(\theta)=
 \frac{1}{2}\left(\begin{array}{cc}
                   1 & 1 \\
                   -1 & 1
                  \end{array}\right)
 \mathcal{G}_\lambda^\prime(\theta)
\left(\begin{array}{cc}
       1 & -1 \\
       1 & 1 
      \end{array}
\right)
\end{equation*}
where
\begin{equation*}
 \mathcal{G}_\lambda^\prime(\theta)=\left(\begin{array}{cc}
        \lambda & {\rm e}^{{\rm i}\xi(\theta)}\\
        {\rm e}^{-{\rm i}\xi(\theta)} & \lambda
       \end{array}\right).
\end{equation*}
and
\begin{equation}\label{xi_log}
 \cos\xi(\theta)=
 \frac{F(\theta)}
 {\sqrt{F(\theta)^2+|G_\nu(\theta)|^2}},
 \quad \sin\xi(\theta)=\frac{{\rm i}G_\nu(\theta)}
 {\sqrt{F(\theta)^2+|G_\nu(\theta)|^2}}.
\end{equation}
The determinants of the block Toeplitz matrices
with $\mathcal{G}_\lambda$ and 
$\mathcal{G}_\lambda^\prime$ are equal and, 
therefore, 
$$\log D_X(\lambda)=
\log D_X[\mathcal{G}_\lambda^\prime].$$
In the following, it will be convenient to work with the symbol
$\mathcal{G}_\lambda^\prime$ instead of 
$\mathcal{G}_\lambda$.

We introduce now the symbol
\begin{equation*}
 \tilde{\mathcal{G}}_\lambda(\theta)
 =
 \left(\begin{array}{cc}
                  \lambda & {\rm e}^{{\rm i}|\xi(\theta)|}\\
                  {\rm e}^{-{\rm i}|\xi(\theta)|} & \lambda
                 \end{array}\right).
\end{equation*}

The behaviour of $|\xi(\theta)|$ 
in the neighbourhood of $\theta=0$ is similar
to that of $|\log g_\nu(\theta)|$. If we calculate
the leading term of the Fourier coefficients 
$\tilde{s}_k^{(\nu)}$ of $|\log g_\nu(\theta)|$ 
we find
\begin{equation*}
 \tilde{s}_k^{(\nu)}=O\left(\frac{1}{k(\log|k|)^{\nu+1}}\right).
\end{equation*}
This means that 
$$\sum_{k=1}^\infty k 
\tilde{s}_k^{(\nu)}\tilde{s}_{-k}^{(\nu)}<\infty$$
for any $\nu>0$ and it is well-behaved in the sense of
(\ref{well-behaviour}) hence the Strong Szeg\H{o} Theorem
(\ref{sst}) is satisfied. Therefore, one may conclude
that the symbol $\tilde{\mathcal{G}}_\lambda$ is
also well-behaved and verifies the Widom theorem for 
block Toeplitz determinants (\ref{widom}),
see the Appendix.

Since $\tilde{\mathcal{G}}_\lambda$ is well-behaved, we 
can apply the Basor localisation theorem (\ref{loc}) 
to the product of symbols $\mathcal{G}_\lambda^\prime 
\tilde{\mathcal{G}}_\lambda^{-1}$. Then we 
have 
\begin{equation*}
 \lim_{|X|\to\infty}\frac{D_X[\mathcal{G}_\lambda^\prime \tilde{\mathcal{G}}_\lambda^{-1}]}
 {D_X[\mathcal{G}_\lambda^\prime]D_X[\tilde{\mathcal{G}}_\lambda^{-1}]}<\infty.
\end{equation*}
This implies that 
\begin{equation*}
 \log D_X(\lambda)=
 \log D_X[\mathcal{G}_\lambda^\prime]=
 \log D_X[\mathcal{G}_\lambda^\prime 
 \tilde{\mathcal{G}}_\lambda^{-1}]-
 \log D_X[\tilde{\mathcal{G}}_\lambda^{-1}]+O(1).
\end{equation*}
Now, we can make use of Widom theorem (\ref{widom}) 
to derive the asymptotic behaviour of $\log D_X[\tilde{\mathcal{G}}_\lambda^{-1}]$.
According to it,
\begin{equation*}
 \log D_X[\tilde{\mathcal{G}}_\lambda^{-1}]=-\log(\lambda^2-1)|X|+O(1)
\end{equation*}
because $\det \tilde{\mathcal{G}}_\lambda^{-1}(\theta)=(\lambda^2-1)^{-1}$.
Therefore,
\begin{equation}\label{step_A}
 \log D_X(\lambda)=
 \log D_X[\mathcal{G}_\lambda^\prime 
 \tilde{\mathcal{G}}_\lambda^{-1}]+\log(\lambda^2-1)|X|+O(1).
\end{equation}

Note that, by the definition of $\tilde{\mathcal{G}}_\lambda$, 
the product $\mathcal{G}_\lambda^\prime 
 \tilde{\mathcal{G}}_\lambda^{-1}$ is
\begin{equation*}
 \mathcal{G}_\lambda^\prime(\theta)
 \tilde{\mathcal{G}}_\lambda(\theta)^{-1}= I,
 \,\,\, \mbox{for}\,\,\,
 -\pi\leq \theta< 0,
\end{equation*}
and
\begin{equation*}
 \mathcal{G}_\lambda^\prime(\theta)
 \tilde{\mathcal{G}}_\lambda(\theta)^{-1}=
 \frac{1}{\lambda^2-1}
 \left(\begin{array}{cc}
  \lambda^2-{\rm e}^{2{\rm i}\xi(\theta)} & 
  2{\rm i}\lambda \sin\xi(\theta) \\
  -2{\rm i}\lambda \sin\xi(\theta) & 
  \lambda^2-{\rm e}^{-2{\rm i}\xi(\theta)}
       \end{array}\right),\,\,\, \mbox{for}
       \,\,\, 0\leq\theta<\pi.
\end{equation*}
Then the unitary matrix 
\begin{equation*}
 U(\theta)=\frac{1}{\sqrt{2}}\left(\begin{array}{cc}
             U_+(\theta) & U_-(\theta) \\
             U_-(\theta) & U_+(\theta)
           \end{array}\right),
\end{equation*}
with
\begin{equation*}
 U_\pm(\theta)=\left(1\pm \sqrt{1-\lambda^2\,
 {\rm sec}\,|\xi(\theta)|}\right)^{1/2},
\end{equation*}
diagonalizes the symbol $\mathcal{G}_\lambda^\prime
\tilde{\mathcal{G}}_\lambda^{-1}$,
\begin{equation*}
 U(\theta)\mathcal{G}_\lambda^\prime \tilde{\mathcal{G}}_\lambda^{-1}
 U(\theta)^{-1}=\left(\begin{array}{cc}
                       \mu_\lambda^-(\theta) & 0 \\
                       0 & \mu_\lambda^+(\theta)
                      \end{array}\right).
\end{equation*}
where the eigenvalues 
are $$
\mu_\lambda^\pm(\theta)=1,\,\,\, \mbox{for}\,\,\, -\pi\leq\theta<0,$$
and
\begin{equation*}
 \mu_\lambda^\pm(\theta)=\left(\frac{\sqrt{\lambda^2-\cos^2\xi(\theta)}
 \pm \sin\xi(\theta)}{\sqrt{\lambda^2-1}}\right)^2,\,\,\, \mbox{for}\,\,\, 
 0\leq \theta<\pi.
\end{equation*}
Since $U(\theta)$ only depends on $|\xi(\theta)|$ we
may conclude that it is well-behaved by the same reasons
as those given for the symbol $\tilde{\mathcal{G}}_\lambda$.
Then we can apply again the localisation theorem (\ref{loc})
to the product $U\mathcal{G}_\lambda^\prime 
\tilde{\mathcal{G}}_\lambda^{-1}U^{-1}$, and
\begin{equation*}
 \log D_X[\mathcal{G}_\lambda^\prime \tilde{\mathcal{G}}_\lambda^{-1}] = 
 \log D_X[U\mathcal{G}_\lambda^\prime \tilde{\mathcal{G}}_\lambda^{-1} U^{-1}]
 -\log D_X[U]-\log D_X[U^{-1}]+O(1).
\end{equation*}
Due to the well-behaviour of $U$, the Widom theorem
(\ref{widom}) gives the asymptotic limit of $\log D_X[U]$ 
and $\log D_X[U^{-1}]$. In fact, since $\det U(\theta)=1$,
the linear, dominant term of their expansion in $|X|$ vanishes
and the rest of the terms tend to a finite value when 
$|X|\to\infty$. This leads us to conclude that
\begin{equation}\label{step_B}
 \log D_X[\mathcal{G}_\lambda^\prime \tilde{\mathcal{G}}_\lambda^{-1}] = 
 \log D_X[U\mathcal{G}_\lambda^\prime \tilde{\mathcal{G}}_\lambda^{-1} U^{-1}]
+O(1).
\end{equation}

Notice that the product $U\mathcal{G}_\lambda^\prime \tilde{\mathcal{G}}_\lambda^{-1} U^{-1}$
is diagonal and, therefore, the corresponding block Toeplitz matrix can be expressed,
after a global change of basis, as
the direct sum of two Toeplitz matrices with symbol the eigenvalues 
$\mu_\lambda^+(\theta)$ and $\mu_\lambda^-(\theta)$. This fact implies that
\begin{equation}\label{asymp_eigen}
 \log D_X[U\mathcal{G}_\lambda^\prime \tilde{\mathcal{G}}
 _\lambda^{-1} U^{-1}]=
 \log D_X[\mu_\lambda^+]+\log D_X[\mu_\lambda^-],
\end{equation}
that allows to apply the conjecture for Toeplitz 
determinants with anomalous symbol that we have 
proposed in Section \ref{sec:sublog_conj}.
According to it the Toeplitz determinants with
symbol $\mu_\lambda^\pm$ behave as
\begin{equation}\label{conj_eigenval}
\log D_X[\mu_\lambda^\pm]=
s_0^\pm(\lambda) |X|+
\sum_{k=1}^{|X|} k s_k^\pm(\lambda)s_{-k}^\pm(\lambda)
+o(1),
\end{equation}
where 
$$s_k^\pm(\lambda)=\frac{1}{2\pi}\int_{-\pi}^\pi
\log \mu_\lambda^\pm(\theta){\rm e}^{{\rm i}\theta k}
{\rm d}\theta
$$
are the Fourier coefficients of the logarithm of
the eigenvalues $\mu_\lambda^\pm(\theta)$.

Since the zero modes satisfy $s_0^+(\lambda)=-s_0^-(\lambda)$ 
the linear terms of the expansion (\ref{conj_eigenval}) 
cancel in (\ref{asymp_eigen}) and, therefore,
\begin{equation}\label{dx_uggu}
 \log D_X[U\mathcal{G}_\lambda^\prime 
 \tilde{\mathcal{G}}_\lambda^{-1} U^{-1}]=
 \sum_{k=1}^{|X|}
 k \left[s_k^+(\lambda) s_{-k}^+(\lambda)+
 s_k^-(\lambda) s_{-k}^-(\lambda)\right]+
 O(1).
\end{equation}

Collecting (\ref{step_A}), (\ref{step_B})
and (\ref{dx_uggu}), we conclude that the 
asymptotic behaviour of $\log D_X(\lambda)$ 
is given by
\begin{equation}\label{dx_log}
 \log D_X(\lambda)=\log(\lambda^2-1)|X|+
 \sum_{k=1}^{|X|}
 k \left[s_k^+(\lambda) s_{-k}^+(\lambda)+
 s_k^-(\lambda) s_{-k}^-(\lambda)\right]+
 O(1).
\end{equation}
Now, if we insert the linear term of this expansion
in the contour integral (\ref{ent_contour}) it vanishes
\footnote{The log derivative in (\ref{ent_contour})
  for the linear term produces simple poles at $\lambda=\pm1$
  and since $f_\alpha(\pm1)=0$ the integral vanishes} 
and, therefore, the leading contribution to 
the entanglement entropy is given by the sum.

Hence to obtain the growth rate of the latter
we need to study the asymptotic behaviour of the
Fourier coefficients, which are given by
\begin{equation*}
 s_k^\pm(\lambda)=
 \frac{1}{\pi}\int_{0}^{\pi} \log
 \frac{\sqrt{\lambda^2-\cos^2\xi(\theta)}
 \pm \sin \xi(\theta)}{\sqrt{\lambda^2-1}}
 {\rm e}^{{\rm i}k\theta}{\rm d}\theta,
\end{equation*}
where $\cos\xi(\theta)$ and $\sin\xi(\theta)$ are as given in  (\ref{xi_log})
and we have used that $\mu_\lambda^\pm(\theta)=1$ for $-\pi\leq \theta<0$. 

Standard estimation techniques, like the change of variable to $k\theta$  and the 
rotation of the integration domain to the imaginary axis, lead to the following
asymptotic behaviour
\begin{equation}\label{sk_eigen2}
 s_k^\pm(\lambda)= \frac{{\rm i}}{k\pi}
 \log \left( 1\mp 
 \frac{\pi/(h+2)}
      {(\log |Ak|)^\nu\sqrt{\lambda^2-1}}\right)+
  O\left(\frac{1}{(\log |k|)^{\nu+1}}\right).
\end{equation}
Now we are equipped with all the ingredients to derive the asymptotic 
behaviour of the R\'enyi entanglement entropy.

First notice that, as it was established before, the linear term vanishes and
the contour integral reduces to
\begin{equation*}
 S_{\alpha,X}=
 \frac{1}{4\pi{\rm i}}
 \lim_{\varepsilon\to1^+}
 \oint_{\cont}
 f_\alpha(\lambda/\varepsilon)
 \frac{{\rm d}}{{\rm d}\lambda}
 E_X(\lambda){\rm d}\lambda,
\end{equation*}
where
\begin{equation}\label{E_X}
 E_X(\lambda)=\sum_{k=1}^{|X|}
 k \left[s_k^+(\lambda) s_{-k}^+(\lambda)+
 s_k^-(\lambda) s_{-k}^-(\lambda)\right]+
 O(1).
\end{equation}

\begin{figure}[H]
  \centering
    \includegraphics[width=0.8\textwidth]{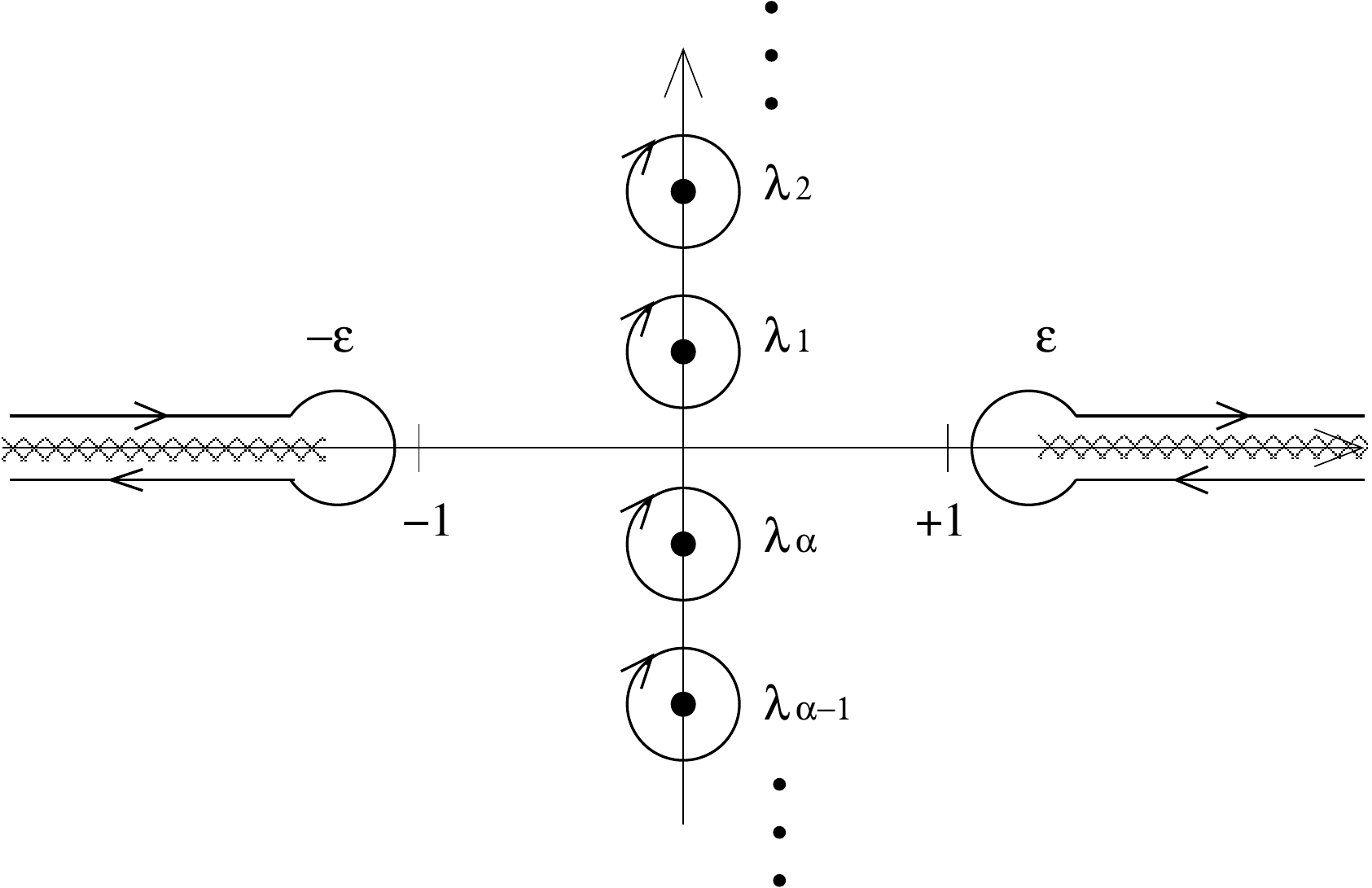}
    \caption{Contours of integration for the computation of the
      R\'enyi entanglement entropy in (\ref{ent_correl_log_parts}).
      They surround clockwise the poles and branch cuts of the function
      ${\rm d}f_\alpha(\lambda/\varepsilon)/{\rm d}\lambda$.
      Poles are the points at the imaginary axis $\lambda_l$
      given by (\ref{poles_f_alpha}) while branch cuts correspond to the
      intervals $(-\infty,-\varepsilon]$ and $[\varepsilon,\infty)$.}
  \label{contorno1}
\end{figure}

Performing an integration 
by parts, one obtains
\begin{equation}\label{ent_correl_log_parts}
 S_{\alpha,X}=-\frac{1}{4\pi {\rm i}}
 \lim_{\varepsilon\to1^+}
 \oint_{\cont}
 \frac{{\rm d}f_\alpha(\lambda/\varepsilon)}
 {{\rm d}\lambda}E_X(\lambda){\rm d}\lambda.
 \end{equation}
Since $s_k^\pm(\lambda)$ is analytic outside the real interval 
$[-1,1]$, we can evaluate the latter integral by deforming 
the integration contour to infinity, surrounding the
poles and cuts of ${\rm d}f_\alpha/{{\rm d}\lambda}$
as it is represented in Fig. \ref{contorno1}.
Inserting in $E_X$ the asymptotic behaviour of
$s_k$ given in (\ref{sk_eigen2}) and replacing
the sum $\displaystyle \sum_{k=1}^{|X|}$ by an
integral $\displaystyle\int_0^{|X|}{\rm d} k$
we can, in principle, obtain the asymptotic behaviour
for $S_{\alpha,X}$.

While the previous computation, as it has been described,
is perfectly feasible, we obtain rather cumbersome expressions
that add little understanding to our study of the sublogarithmic
law for the entanglement entropy. However, when $\alpha$ is an
integer greater than 1 the problem simplifies notably and we
are able to give very explicit formulae.

In this case the cuts in Fig. \ref{contorno1} disappear and the
only singularities of ${\rm d}f_\alpha(\lambda)/{\rm d}\lambda$ 
are simple poles at the points of the imaginary axis
\begin{equation}\label{poles_f_alpha}
 \lambda_l={\rm i}\tan\frac{\pi(2l-1)}{2\alpha},
 \quad l=1,\dots,\alpha,\quad l\neq \frac{\alpha+1}{2}.
\end{equation}
Therefore, $S_{\alpha,X}$ can be computed from the sum of
the residues of the integrand in (\ref{ent_correl_log_parts})
at these points,
\begin{equation}\label{renyi_ent_log}
 S_{\alpha,X}=\frac{1}{2(1-\alpha)}
 \sum_{l=1}^\alpha E_X(\lambda_l).
\end{equation}
Using the asymptotic behaviour of $s^\pm_k$ given in (\ref{sk_eigen2})
and restricting to first order we have  
\begin{eqnarray*}
 ks_k^\pm(\lambda_l) 
 s_{-k}^\pm(\lambda_l)=-\frac{c(\lambda_l)}{k(\log|Ak|)^{2\nu}}
 +O\left(\frac{1}{k(\log|k|)^{3\nu}}\right),
\end{eqnarray*}
where
\begin{equation*}
 c(\lambda)=\frac{1}{(h+2)^2(|\lambda|^2+1)}.
\end{equation*}

If we plug this into (\ref{renyi_ent_log}) and approximate
the sum over $k$ by an integral,
we obtain that the R\'enyi entanglement entropy for integer
$\alpha>1$ grows with $|X|$ as
\begin{equation}\label{renyi_expan_log}
 S_{\alpha,X}=C_\alpha^{(\nu)} (\log |X|)^{1-2\nu}+
 O\left((\log|X|)^{1-3\nu} + 1\right),
 \quad \mbox{for}\quad  0<\nu<1/2,
\end{equation}
where
\begin{eqnarray}\label{coeff_sublog}
C_\alpha^{(\nu)}&=&\frac{1}{\alpha-1}\sum_{l=1}^\alpha
\frac{c(\lambda_l)}{1-2\nu}\cr
&=&\frac\alpha{2(\alpha-1)}\frac1{(h+2)^2(1-2\nu)}.
\end{eqnarray}
On the other hand, 
\begin{equation*}
 S_{\alpha,X}=C_\alpha^{(1/2)} \log\log |X|+
 O\left(1\right),
 \quad \mbox{for}\quad  \nu=1/2,
\end{equation*}
with
\begin{eqnarray}\label{coeff_loglog}
  C_\alpha^{(1/2)}&=&\frac{1}{\alpha-1}\sum_{l=1}^\alpha c(\lambda_l)\cr
&=&\frac\alpha{2(\alpha-1)}\frac1{(h+2)^2}.
\end{eqnarray}
In conclusion, we expect a sublogarithmic growth
of the R\'enyi entanglement entropy in the ground state of
the Hamiltonian (\ref{ham_log}) with logarithmic decaying 
pairings. To our knowledge, 
this kind of behaviour has not been reported in the literature.
A striking feature of our result is that 
the next terms in the expansion (\ref{renyi_expan_log}) are of the form
$(\log|X|)^{1-m\nu}$ with $m=3,4,\dots$. Therefore, 
all the terms for which $m<1/\nu$ also diverge in the 
limit $|X|\to\infty$. This makes that very similar, divergent
terms mix up and, therefore, it is difficult to isolate 
numerically the leading asymptotic behaviour.
{\color{black}  Hence, in order to test our predictions
we must allow for subdominat terms.

In Fig. \ref{sublog_entropy}
we represent the R\'enyi entanglement entropy for
$\alpha=2$ with $\nu=0.25$ (squares) and $\nu=0.5$ (circles),
taking in both cases $h=1/2$ and $A=2$, as a function of the size of the subsystem $|X|$.
The curves are  obtained by supplementing the predicted
coefficient for the leading asymptotic term of (\ref{coeff_sublog}) and
(\ref{coeff_loglog}) with the next two subleading
corrections. The values of the latter coefficients have been computed by
fitting the curve to the numerical results in the interval between $|X|=100$
to $|X|=1000$.

The continuous line, associated to $\nu=0.25$ represents the function
$$F_{0.25}(x)= \frac8{25}(\log x)^{\frac12} + a (\log x)^{\frac14} + b, \quad \mbox{with}\quad a=-0.400,\  b=0.811.$$

The dot-dashed line corresponds to $\nu=0.5$ and is given by
$$F_{0.50}(x)= \frac4{25}\log\log x+ c+ d (\log x)^{-\frac12} , \quad \mbox{with}\quad c=0.474,\  d=0.256.$$

Note that the curves are very close to the numerical results.
This fact is specially noticeable if we consider that the fit for
$a, b, c$ and $d$ has been performed in the window between $100$
and $1000$ and then we extend the curve
up to $|X|=10\,000$. We believe that this extraordinary agreement is a
strong support for the validity of our results.

\begin{figure}[H]
  \centering
    \includegraphics[width=0.9\textwidth]{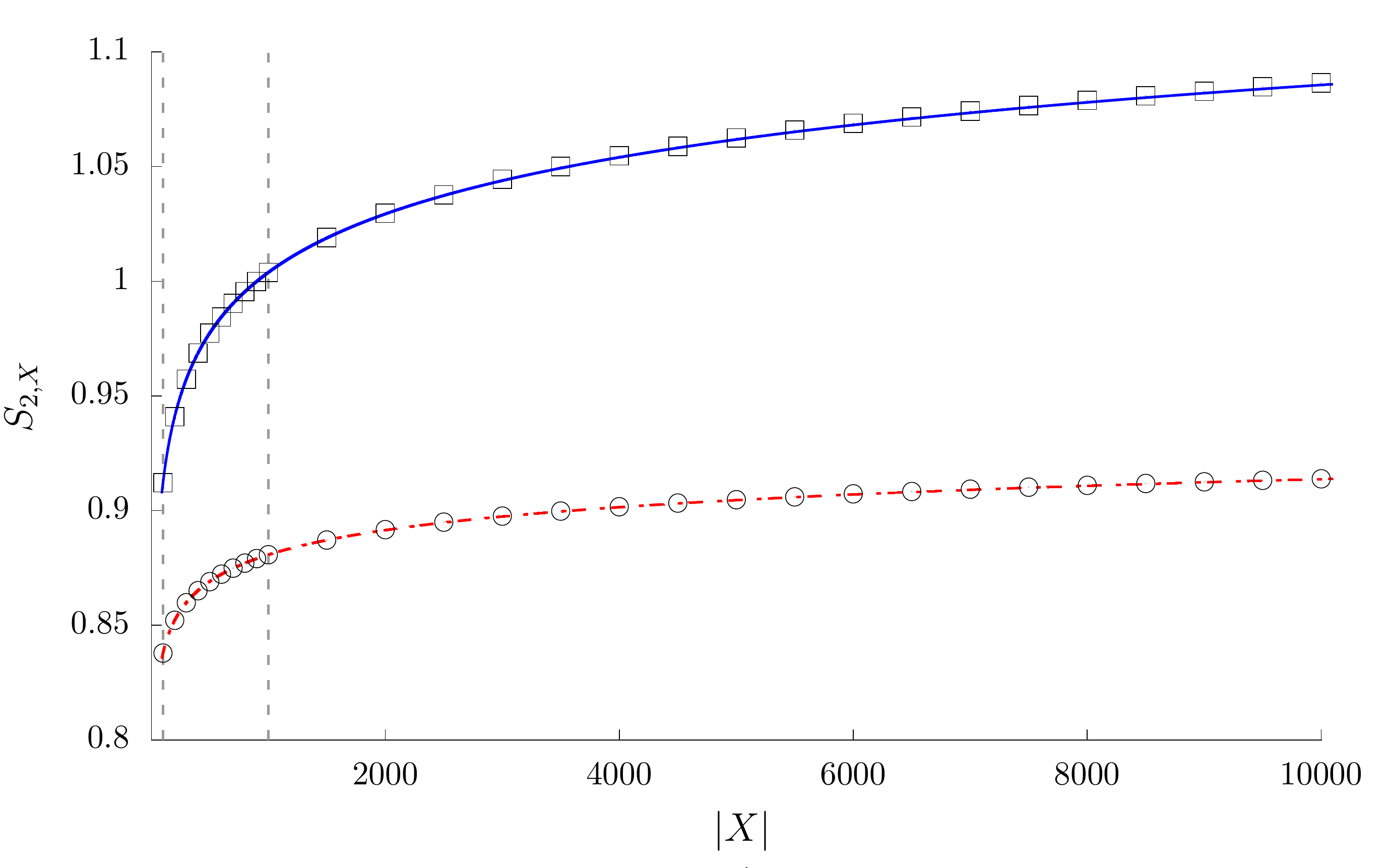}
    \caption{\color{black}The squares and circles represent the numerical value of the
      R\'enyi entanglement entropy for $\alpha=2$ with $\nu=0.25$
      and $\nu=0.5$ respectively as a function of $|X|$.
      The continuous line contains our prediction for the leading term
      for $\nu=0.25$ supplemented with the two next subleading corrections
      $F_{0.25}(x)= \frac8{25}(\log x)^{\frac12} -
      0.400 (\log x)^{\frac14} + 0.811$. The dot-dashed line is the analogous
      for $\nu=0.5$, i. e. $F_{0.50}(x)= \frac4{25}\log\log x+0.474 + 0.256 (\log x)^{-\frac12}$.
    }
      \label{sublog_entropy}
\end{figure}
}

\section{Conclusions}\label{sec:conclusions}
In this paper we have investigated the possibility of unexplored
behaviours for the R\'enyi entanglement entropy in fermionic chains,
different from the linear and the logarithmic growth or the
saturation to a constant value with the length of the interval.

Our investigation is based on the relation between block Toeplitz
determinants and the R\'enyi entanglement entropy in the ground
state of quadratic, homogeneous fermionic chains. In particular, we
have considered chains where the pairing couplings decay logarithmically
with the distance. For these systems, the ground state entanglement
entropy is related to the Toeplitz determinant generated by a family
of continuous, non-zero symbols that do not lie within the domain of
validity of the Strong Szeg\H{o} theorem.

We have proposed a new asymptotic regime for these symbols that
leads to the sublogarithmic growth of the logarithm of the Toeplitz
determinant and, therefore, of the R\'enyi entanglement entropy.
That is, the leading term of the ground state entanglement entropy
in these fermionic chains is $(\log |X|)^a$ for some $0<a<1$ related to the
dumping exponent in the pairing couplings, and $|X|$ the length
of the interval in the chain. We have also found a case with a
$\log\log|X|$ behaviour.

The numerical tests performed support the conjecture
for this type of Toeplitz determinants.

It is remarkable that this asymptotic regime both
for Toeplitz determinants and for the entanglement entropy
has not been previously reported in the literature. In
the logarithmic conformal field theories studied in \cite{Bianchini},
they found terms of the form $\log \log|X|$ but, contrary
to our case, they are subleading,
\textcolor{black}
{see also \cite{Couvreur,Sierra3}}. The same occurs with the
sublogarithmic terms detected numerically in \cite{Koffel} in long range
quantum Ising chains. It would be interesting to understand
the leading sublogarithmic terms that we have found here using
field theory methods as well as to investigate if they appear in
other models{\color{black}, and in particular, whether there is any direct relation 
between our results for $\nu=1/2$ and some logarithmic CFT}.
\newline

\textbf{Acknowledgments:} Research partially supported by grants
E21$\_$17R, DGIID-DGA and FPA2015-65745-P, MINECO (Spain). FA was
supported by FPI Grant No. C070/2014, DGIID-DGA/European Social Fund
and by Brazilian Ministries MEC and MCTIC.
ZZ was supported by the Janos Bolyai Scholarship, and the National
Research Development and Innovation Office of Hungary under Project
No. 2017-1.2.1-NKP-2017-00001, and by  Grants No. K124152 and KH129601.
The authors thankfully acknowledge the resources from the supercomputer
\textit{Memento}, technical expertise and assistance provided by BIFI
(Universidad de Zaragoza).

\appendix
\section{}
 
 In this Appendix we will present several results on, so called,
 block Toeplitz determinants that are necessary for the proper
 understanding of the paper.

 A block Toeplitz matrix with $d$-dimensional matrix valued symbol
 $J:S^1 \rightarrow M_d(\mathbb{C})$ is a $(N\!\cdot\! d)\times (N\!\cdot\! d)$
 matrix formed by $d$ dimensional blocks which are given by the Fourier
 coefficients of $J$, namely
\begin{equation*}
  (T_N[J])_{nm}=J_{n-m}=\frac{1}{2\pi}\int_{-\pi}^\pi
  J(\theta) {\rm e}^{{\rm i}\theta(n-m)}{\rm d}\theta,\quad n,m=1,\dots,N.
\end{equation*}

We will be mainly interested in the determinant of this matrix that
we will denote by $D_N[J]=\det (T_N[J])$.

Gyires \cite{Gyires} found a generalisation of the Szeg\H{o} theorem 
(\ref{szego}) for the determinant of block Toeplitz matrices that was 
later extended by Hirschman \cite{Hirschman} and Widom \cite{Widom} to 
a wider variety of symbols. According to this, the leading term in 
the asymptotic expansion of $\log D_N[J]$ should be also linear 
\begin{equation}\label{widom0}
  \log D_N[J]=\frac{N}{2\pi}\int_{-\pi}^\pi
  \log\det J(\theta){\rm d}\theta+o(N),
\end{equation}
provided $\det J(\theta)\neq 0$ and
the argument of $\det J(\theta)$ is continuous 
and periodic for $\theta\in[-\pi, \pi]$ (zero 
winding number). 

Analogously to the scalar case, if in addition the symbol
$J(\theta)$ is well-behaved  in the sense that
\begin{equation*}
  \sum_{k=-\infty}^\infty \norm{J_k}
  +\sum_{k=-\infty}^\infty|k| \norm{J_k}^2<\infty,
\end{equation*}
where $\norm{\cdot}$ is the Hilbert-Schmidt norm
of the $d\times d$ matrices, the \textbf{Widom theorem} 
\cite{Widom, Widom2, Widom3} states that the next contribution 
in the expansion (\ref{widom0}) should be finite when $|X|\to\infty$,
\begin{equation}\label{widom}
  \log D_N[J]=\frac{N}{2\pi}\int_{-\pi}^\pi
  \log\det J(\theta){\rm d}\theta+E[J]+o(1).
\end{equation}
If we call $T[J]$ the semi-infinite matrix obtained
from $T_N[J]$ when $N\to\infty$, the constant term $E[J]$
reads
\begin{equation}\label{e_widom}
 E[J]=\log \det T[J] T[J^{-1}].
\end{equation}
The Widom theorem reduces to the Strong Szeg\H{o} theorem (\ref{sst}) when
the symbol is a scalar function, i.e. $d=1$.
This result will be important for us in order to find the asymptotic behaviour
of the entanglement entropy for the long range Kitaev chain (\ref{ham_log}).

Another theorem that we will be needed troughout the paper is
the \textbf{localisation theorem} found by E. Basor in \cite{Basor}.
It reads as follows:

 If we consider two $d\times d$ symbols
$J_1(\theta)$, $J_2(\theta)$ such that their block Toeplitz matrices
$T_X[J_1]$ and $T_X[J_2]$ are invertible for $|X|$ large enough and
the semi-infinite matrices $T[J_1J_2]-T[J_1]T[J_2]$, $T[J_2J_1]-T[J_2]T[J_1]$ 
are trace-class then
\begin{equation}\label{loc}
  \lim_{|X|\to\infty}\frac{D_{X}[J_1J_2]}{D_{X}[J_1]D_{X}[J_2]}<\infty. 
\end{equation}

The operator $T[J_1J_2]-T[J_1]T[J_2]$ is trace-class if there exists
a smooth partition of the unit $\{f_1(\theta), f_2(\theta)\}$ such that the
derivatives of $J_1f_1$ and $J_2f_2$ are H\"older continuous with an exponent
greater than $1/2$. Or, in more informally words, the operator
$T[J_1J_2]-T[J_1]T[J_2]$ is trace-class provided $J_1$ and $J_2$
are not \textit{bad} (non smooth) at the same point.

The localisation theorem implies that if the symbol $J(\theta)$ 
has several discontinuities at $\theta_1$, $\dots$, $\theta_R$, the
divergent contribution in the large $|X|$ limit of each one to $D_X[J]$ is
independent from the rest and they can be studied separately.

Although it is not really necessary in this paper,
for the sake of completeness we would like to mention
one more result found in \cite{Ares6} that allows
to compute the logarithmic divergence of block Toeplitz
determinants with symbols that have jump discontinuities.

This is a generalization of the scalar case mentioned
in Section \ref{sec:sublog_conj} and can be stated as
follows:

Let $J$ be a piecewise smooth $d \times d$
matrix valued symbol such that $\det J(\theta)\neq 0$ and has
zero winding number. If $J$ has jump discontinuities
at $\theta_1$, $\dots$, $\theta_R$, 
where $J_\r^\pm$ are the lateral limits of $J(\theta)$ at $\theta_\r$,
the determinant 
$D_X[J]$ of its block Toeplitz matrix is
\begin{equation}\label{log_det}
  \log D_X[J]=\frac{|X|}{2\pi}\int_{-\pi}^\pi \log\det J(\theta){\rm d}\theta+
  \frac{\log|X|}{4\pi^2}\sum_{\r=1}^R \Tr\left(\log\left( J_\r^-(J_\r^+)^{-1}\right)^2\right)+O(1).
\end{equation}

 \end{document}